\documentclass{aa}

\usepackage{graphicx}
\usepackage{txfonts}

\usepackage[usenames,dvipsnames]{color}
\usepackage[switch]{lineno}
\usepackage[]{hyperref}

\begin{document}

   \title{Hadronic emission from the environment of the Crab Pulsar Wind Nebula by re-accelerated particles}
    \titlerunning{Hadronic emission from the Crab environment by re-accelerated particles}

   \author{S.T. Spencer
          \inst{1,2},
          A.M.W. Mitchell\inst{1}
          \and
          B. Reville\inst{3}
          }
    \authorrunning{S.T. Spencer et al.}
   \institute{Erlangen Centre for Astroparticle Physics (ECAP), Friedrich-Alexander-Universit{\"a}t Erlangen-N{\"u}rnberg, Nikolaus-Fiebiger-Str. 2, D 91058 Erlangen, Germany\\
              \email{samuel.spencer@fau.de}
         \and
             Department of Physics, Clarendon Laboratory, Parks Road, Oxford, OX1 3PU, United Kingdom
         \and
             Max-Planck-Institut f{\"ur} Kernphysik, Saupfercheckweg 1, 69117 Heidelberg, Germany
             }

   \date{Received September 15, 2001; accepted September 16, 3001}

  \abstract 
   {The observation of peta-electronvolt (PeV) $\gamma$-ray photons from the Crab Nebula by LHAASO has revitalised the possibility of a secondary population of hadrons producing the highest energy emission through neutral pion decay. Despite previous studies modelling this population, the origin of such high-energy hadronic particles remains unclear.}
   {We consider possible acceleration scenarios for multi PeV particles in the Crab Nebula environment, including one in which high-energy protons produced at the supernova remnant's outer shock diffuse into the pulsar wind nebula. Particles which reach the Crab Pulsar's wind termination shock can be accelerated to the required energies, and subsequently interact with the dense filaments surrounding the nebula.}
   {We perform particle transport simulations of this scenario, including the effects of the expansion of the pulsar wind nebula into the surrounding supernova ejecta.}
   {We find that this results in PeV photons being produced over the lifetime of the Crab system, without over-estimating the flux at lower energies or exceeding the energy budget of the Crab Pulsar. This results in a reasonable match to the LHAASO data at the highest energies. We also present predictions for the resulting all-flavour neutrino flux, finding it to be approximately an order of magnitude below the sensitivity of current generation instruments.
 }
   {}


   \maketitle


\section{Introduction}

The Crab Nebula is the remnant of a star purported to have exploded in 1054 AD 
\citep{Lundmark_1921}. It is the most widely studied object in Very-High-Energy (VHE) $\gamma$-ray astrophysics, and was the first (and brightest persistent) TeV $\gamma$-ray source detected from the ground using the Imaging Atmospheric Cherenkov Telescope (IACT) technique \cite[e.g.][]{1989whipple,Hegra,MagicCrab,hesscrab1}. More recently, the $\gamma$-ray emission from the Crab Nebula has been found to have an extension of $(52.2\pm2.9_{\rm stat}\pm6.6_{\rm sys})''$ at TeV energies in a H.E.S.S.-only analysis \citep{crabextension}, and $r_{68}=(0.82\pm0.29_{\rm stat})'$ in a joint Fermi-LAT and H.E.S.S. analysis \citep{aharonian2024spectrum}. The latter analysis found strong evidence for a decreasing size as a function of energy in the $\gamma$-ray regime up to $\sim 100\,\mathrm{TeV}$; the energy-dependent morphology in this range is likely lepton-dominated. Furthermore, pulsed emission from the central pulsar, PSR J0534+2200, has also been detected at TeV photon energies by MAGIC \citep{magiccrabpulsar} and VERITAS \citep{veritascrabpulsar}. This pulsar is located at a distance of 2\,kpc from Earth \citet{Kaplan_2008}, and has a spin down luminosity of $5 \times 10^{38}\,\mathrm{erg\,s^{-1}}$ \citet{2023ApJ...958..191S}.

As pointed out by \citet{atoyan1996}, production of the TeV $\gamma$-ray emission purely via $\pi^0$ decay can be ruled out on energetic grounds. 
Thus, the primary emission mechanism for $\gamma$-rays above $100\,\mathrm{GeV}$ from the nebula is generally accepted to be Inverse Compton (IC) scattering of photons in the nebula by energetic electrons and/or positrons. 
The target photons come from a variety of background radiation fields, though at photon energies above $\mathrm{100\,TeV}$, the Cosmic Microwave Background (CMB) dominates \citep{lhassocrab}.
The wind termination shock (WTS) of the ultra-relativistic wind of the pulsar, located at $\approx 0.13\,\mathrm{pc}$ from the pulsar \citep{Weisskopf2012}, is thought to be the site of acceleration of the TeV emitting particles \cite[e.g.][]{reesgunn, Bell92, amato2006,Giacinti_2018}, though the details of the acceleration remain unclear. The magnetic field structure of the Crab Nebula has also recently been examined by the Imaging X-ray Polarimetry Explorer (IXPE) mission, revealing a predominantly toroidal magnetic field \citep{ixpe}, as expected for an oblique rotating pulsar \citep[e.g.][]{Coroniti, Michel, Bogovalov}.

Despite a purely hadronic origin for the GeV-TeV emission being ruled out, \cite{atoyan1996} proposed that a hard component could in principle dominate at higher energies. 
Such a scenario becomes increasingly favourable at ultra-high energies, where Klein-Nishina suppression will affect even CMB photons \cite[e.g.][]{dirsonhorns}.
The recent detection of PeV $\gamma$-ray emission by the Large High-Altitude Air Shower Observatory (LHAASO) \citep{lhassocrab} has re-opened this debate. 

Previous studies exploring the role of hadronic emission in the context of the LHAASO results \cite[e.g.][]{LiuWang,peng_multiband_2022,nie}, do not address the origin of the multi-PeV particles required. 
However, even the processes underlying the energisation of inverse-Compton emitting particles are uncertain. It is broadly accepted that the pulsar wind termination shock plays the dominant role, though how exactly remains a topic of active investigation. 
Different models proposed include for example resonant cyclotron absorption of the pairs \cite[e.g][]{amato2003,amato2006} or directly via relativistic shock acceleration \citep{Giacinti_2018, Giacinti_ICRC}. 

The resonant cyclotron absorption mechanism does not accelerate ions efficiently, but in order to produce the observed energetic pairs, it would require the pulsar wind to be energetically dominated by a cold ion component with a large bulk Lorentz factor $\Gamma_{\rm wind} > 10^6$. Protons/ions embedded in such a wind, should they exist, would be thermalised at the shock and injected into the nebula at multi-PeV energies. Despite observational arguments supporting large bulk Lorentz factors \cite[e.g.][]{KC84b}, whether they can be realised in practise remains uncertain \cite[see][for a review]{Kirketal}\footnote{ see also \citet{KirkGiacinti} for a study of the impact of protons on the Crab pulsar wind.}. If the ions are not energetically dominant, or the wind has a substantially lower bulk Lorentz factor, one could consider if relativistic shock acceleration can occur at the termination shock. \citet{Giacinti_2018} have shown that particles of one charge can undergo multiple Fermi cycles at the shock in the equatorial region due to the field reversal across the equator. Both scenarios would require a non-negligible proton/nucleon component in the wind, whose presence is currently unconstrained \cite[though see for example][]{bp97, nie}.

Alternative routes for PeV proton acceleration include diffusive shock acceleration at the outer blast wave of the supernova remnant associated to the progenitor of the Crab pulsar though favourable wind conditions of the progenitor star are needed for particles to exceed PeV energies \citep{Belletal13}. As pointed out by \cite{Bell92}, lower energy protons accelerated at the outer shock can cross the nebula via a grad $B$ drift, and interact with the relativistic termination shock of the pulsar wind \cite[see also][]{LucekBell, BellLucek, ohira}  

\begin{figure}
        \centering
	\includegraphics[width=0.6\columnwidth]{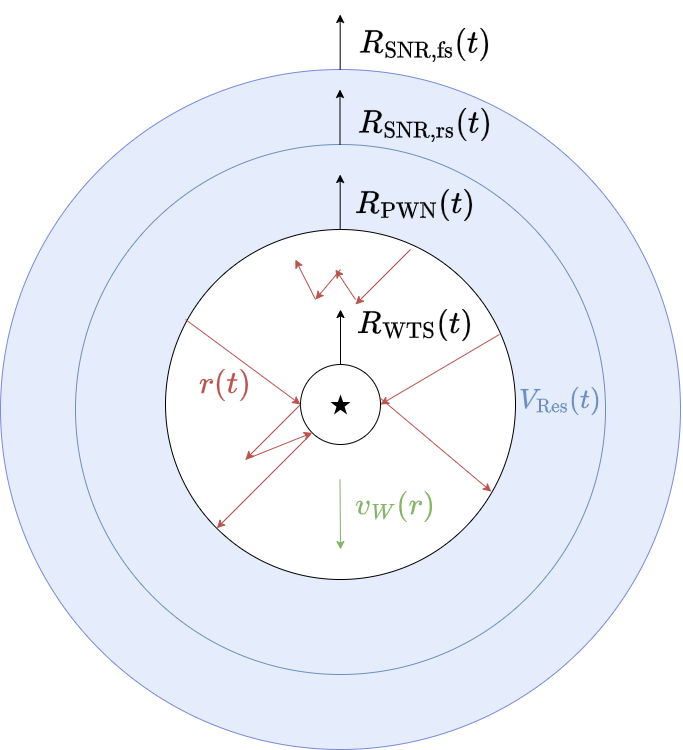}
    \caption{Schematic of our proposed scenario, hadronic particles with position $r(t)$ (shown in red) are initially shocked by the SNR forward shock $R_{\mathrm{SNR,fs}}$. They then travel from the reservoir region (volume $V_{\mathrm{Res}}$, shown in blue) between the $R_{\mathrm{SNR,fs}}$ and the pulsar wind nebula radius to the pulsar wind termination shock $R_{\mathrm{WTS}}$. Following re-acceleration, they advect outwards with the pulsar wind (velocity $v_W$, shown by the green arrow) to hit the reservoir region they originated from. The black star denotes the pulsar's position. For a system the age of the Crab, the supernova reverse shock $R_{\mathrm{SNR,rs}}$ is still travelling outwards.}
    \label{fig:schematic}
\end{figure}

Global magneto-hydrodynamic (MHD) simulations of the Crab system predict a complex magnetic field topology in the enclosed PWN \citep{PorthCrab3D}, through which particles can diffuse against the outward directed flow. We consider the possibility that protons accelerated earlier in the evolution of the forward shock of the SNR populate a reservoir of energetic particles in the region between the PWN boundary and the supernova forward shock. Following \citet{Bell92}, we explore the possibility that particles of sufficiently high energy can in principle penetrate the nebula and reach the WTS. We apply a simple radial diffusion model, and do not consider the role of drift motions. There, protons can be re-accelerated, before escaping the nebula and subsequently re-interacting with gas in the reservoir. A schematic of our proposed scenario is shown in Figure \ref{fig:schematic}. The existence of finger-like structures in the PWN seen in the infrared, believed to be caused by Rayleigh-Taylor instabilities at the PWN boundary, supports this hypothesis. Their presence suggests material is encroaching into the PWN from the surrounding SNR \citep{hester}. The goal of this work is to determine if this proposed scenario can account for the highest energy LHAASO $\gamma$-ray flux points reaching $\gtrsim 1$\,PeV. 

The outline of the remainder of this paper is as follows: in Section \ref{sec:methods} we first make analytic estimates of the diffusion coefficient in the nebula and the resulting $\gamma$-ray flux. We then introduce the theory of an expanding PWN before introducing the Monte-Carlo simulation procedure we adopt and our modelling of the resulting $\gamma$-ray emission. In Section \ref{sec:results} we compare the results of this modelling to the LHAASO data and to the available energy budget from the pulsar, before also making an estimate of the resulting neutrino flux. In Section \ref{sec:conclusions} we present our conclusions.

\section{Methods}
\label{sec:methods}

\subsection{Motivation and estimates}

Working with the hypothesis that the $\gamma$-ray flux at and above PeV energies is dominated by hadronic emission, the LHAASO flux of $\phi(E_\gamma> 10^{15} {\rm~eV}) \gtrsim 10^{-14}$ $\mathrm{erg/cm^2/s}$ can be used to constrain the total energy of protons and other nuclei to be $w_{\rm tot}(E>{\rm PeV})\approx 10^{46}/\langle n_H\rangle$ erg, irrespective of the acceleration process. Here $\langle n_H\rangle$ is the average gas target density surrounding the nebula. This can be compared against the kinetic energy provided by the supernova of $E_{\rm SNR} \approx 10^{51}$ erg, or the integrated energy released by the central pulsar $E_{\rm pulsar} \gg 10^{49}$ erg. The latter inequality assumes the spin-down power was larger in the past. It is evident that a modest efficiency of less than a fraction of a percent would be sufficient in either scenario to account energetically for the required PeV protons. 

Modelling the multi-wavelength observations of the Crab PWN generally indicate that the root mean square field inside the nebula is $B_{\rm PWN} \approx 100~\mu$G. The transport of energetic particles in this field is sensitive to both its global topology and the details of the MHD turbulence embedded in the flow. Based on recent X-ray polarisation measurements \citep{ixpe}, it appears that the radial component of the field is sub-dominant in the nebula and thus the radial diffusion coefficient should not exceed the Bohm limit. Adopting present-day parameters for the nebula, one can estimate an upper limit on the radial diffusion length of $r_{\rm diff}({\rm PeV}) < \sqrt{4 \kappa_{\rm Bohm} t} \approx 2 B_{100\mu{\rm G}}^{-1/2}\,t_{\rm kyr}^{1/2}$\,pc, i.e. comparable to the expected present-day radius of the nebula. This is consistent with the results of 3D MHD simulations of radial diffusion in the Crab Nebula by \citet{PorthDiff}, who found an almost energy independent radial diffusion coefficient of $\lesssim 10^{27}\,\mathrm{cm^2\,s^{-1}}$ for test particles with Lorentz factors $\gamma \ll 10^{10}$. For such small radial diffusion, ions that recently interacted with the wind termination shock remain inside the nebula, while those entering the nebula from outside at the current epoch have a negligible probability of reaching the wind termination shock through diffusion alone \citep[in this work, we do not consider drift motions as done in ][]{Bell92}. Knowledge of the time history and internal structure of the supernova remnant are therefore important for hadronic models of the PeV $\gamma$-ray emission.   

As mentioned in the introduction, we consider the possibility that, during the early stages of the SNR expansion, a fraction of the energy processed by the forward shock was converted to cosmic rays. The particles that accumulate downstream of the external shock enclose the pulsar wind nebula, forming a reservoir of relativistic ions which for simplicity we assume to be comprised exclusively of protons. According to the standard CR-SNR origins model \cite[e.g.][]{Ginzburg} approximately $10\%$ of the SNR's kinetic energy should be converted to CRs over its lifetime. Given the relatively young age of the Crab pulsar, we do not expect the total energy in the reservoir to exceed this. 

It is possible to constrain the total energy, by requiring that emission from the reservoir does not dominate over the inverse-Compton emission from the nebula, at sub PeV energies. This is also a consistency check of our proposed model. Assuming the non-thermal proton distribution in the reservoir is a power-law, $dN/dE \propto E^{-s}$ above 1 GeV, the $\gamma$-ray flux at earth is approximately

\begin{align*}
&\Phi = E_\gamma^2\frac{dN_\gamma}{dE_\gamma dt dA}\\
&\approx 10^{-12} \left(\frac{D}{2~{\rm kpc}}\right)^{-2}\left(\frac{\eta}{10^{-4}}\right)\left(\frac{E_{\mathrm{SN}}}{10^{51}}\right)
\left(\frac{n_H}{10~{\rm cm}^{-3}}\right)\left(\frac{E_\gamma}{\rm GeV}\right)^{2-s} ~\frac{\rm erg}{\rm cm^{2} s}\,,
\end{align*}
where we introduce the efficiency parameter $\eta$ as the fraction of the total  kinetic energy of the SNR in CRs above a GeV, $w_{\rm tot}(E>{\rm GeV}) = \eta E_{\rm SN}$.
For the above numerical values, assuming $s\gtrsim2$ and that the cut-off in the proton distribution extends well beyond several PeV, this alone could accommodate the LHAASO measurements. While this cannot be ruled out based on current observational constraints, it requires that the SNR shock both accelerates and confines particles well beyond a PeV inside the remnant, which is a challenge for current theoretical models \citep{Belletal13}. 

If on the other hand protons accelerated at the SNR's external shock reach some maximum energy $\ll$ PeV, these CRs could fill the volume between the PWN and the outer SNR shock with the desired reservoir. The resulting hadronic emission from these particles would cut off below the LHAASO energy range, with flux level negligible compared to the inverse Compton emission. Similar arguments can be used to rule out a dominant leptonic contribution from the SNR.

\subsection{PWN evolution}

For the spherically symmetric evolution of the system, we follow the approach of \citet{swaluw}. In their model, the PWN radius satisfies
\begin{equation}
R_{\mathrm{PWN}}(t)=1.04 \left(\frac{\dot{E}t_{\mathrm{S}}}{E_{\mathrm{SN}}}\right)^{1/5}\left(\frac{t}{t_{\mathrm{S}}}\right)^{6/5} R_{\mathrm{S}},
\end{equation}
where, $R_{\mathrm{S}}$ is the Sedov radius given by
\begin{equation}
R_{\mathrm{S}}=0.805 t_{\mathrm{S}} \left(\frac{10E_{\mathrm{SN}}}{M_{\mathrm{ej}}}\right)\,,
\end{equation}
and $E_{\mathrm{SN}}=10^{51}\,\mathrm{erg}$ is the supernova energy and $M_{\mathrm{ej}}=3M_{\odot}$ is the supernova ejecta mass. Assuming an ISM density of $n=\mathrm{0.1\,cm^{-3}}$, $t_{\mathrm{S}}$ is the Sedov time given by $t_{\mathrm{S}} \approx \mathrm{10^3\,yr}$ in this case. The velocity of the outer radius of the PWN, $v_{\mathrm{PWN}}$, as a function of time is given by $dR_{\mathrm{PWN}}/dt$,
\begin{equation}
v_{\mathrm{PWN}}=1.04 \times\frac{6}{5}\left(\frac{\dot{E}t_{\mathrm{S}}}{E_{\mathrm{SN}}}\right)^{1/5} \left(\frac{R_{\mathrm{S}}}{t_{\mathrm{S}}}\right)\left(\frac{t}{t_{\mathrm{S}}}\right)^{1/5}
\end{equation}
for the current stage of the Crab's evolution. The WTS radius is fixed as $5\%$ of the PWN radius at each timestep such that the size of the WTS at the current time is approximately correct ($0.1\,\mathrm{pc}$).

\subsection{Simulation procedure}
\label{sec:maths}

We seek to determine what fraction of the particles in the CR reservoir can interact with the WTS, and be accelerated there to higher energies. This requires a number of assumptions on the nature of the particle transport in the nebula, and the acceleration process itself. Analytic treatments of particle acceleration at relativistic shocks is complicated by the fact that the ratio of the shock velocity to the speed of light is not a small parameter that can be exploited \cite[see for example][]{KirkRevilleHuang}. To proceed we make a number of simplifying assumptions. We assume the system is spherically symmetric, and that it can be approximately treated with the non-relativistic transport equation:
\begin{align}
\frac{\partial f}{\partial t}+u_r\frac{\partial f}{\partial r} = \frac{1}{r^2}\frac{\partial }{\partial r}\left[ {r^2 \kappa}\frac{\partial f}{\partial r}\right] + \frac{1}{3r^2} \frac{\partial r^2 u_r}{\partial r} p \frac{\partial f}{\partial p}\,,
\label{trans1}
\end{align}
where $f(r, p, t)$ is the isotropic part of the distribution function, $u_r(r)$ the radial fluid velocity, and $\kappa$ the radial diffusion coefficient.
In general, equation (\ref{trans1}) requires that the distribution remains close to isotropic, and the flow satisfies $u_r \ll c$, both of which break down in the vicinity of the relativistic shock. We discuss how we avoid this complication below. 

 Defining $F=4 \pi p^3 4 \pi r^2 f$, Eq. (\ref{trans1}) can be recast in the form:
\begin{align}
 \frac{\partial F}{\partial t} + \frac{\partial \,}{\partial r}\left[u_{\rm eff} F - \frac{\partial \,}{\partial r}(\kappa F)
 \right] -
\frac{\partial \,}{\partial y}\left[\frac{1}{r^2} \frac{\partial r^2 u_r}{\partial r} \, F\right]=0
 \label{trans2}
\end{align}
where $u_{\rm eff}
(r)= u_r + \frac{1}{r^2}\frac{\partial (r^2 \kappa)}{\partial r}$ is the effective radial velocity and \(y = \ln(E/E_0)\). We choose $E_0=1$\,GeV. For simplicity, we further assume $\kappa$ is constant and uniform in the nebula, and has Bohm scaling 
\begin{equation}
\kappa = \frac{1}{3}\beta r_g c = \beta\left(\frac{Ec}{3eB_{\mathrm{max}}}\right)\,,
\label{eq:kappa}
\end{equation}
where $r_g$ is the particle gyroradius, $E$ its energy, $B_{\mathrm{max}}$ is the magnetic field strength and $\beta$ a free parameter which we set to unity. We set the magnetic field strength to be $112\,\mathrm{\mu G}$ in order to reproduce the synchrotron and IC emission from X-ray wavelengths to PeV at the current epoch with single-zone models \citep{lhassocrab}.

Equation (\ref{trans2}) is of the standard Fokker-Planck form, which readily lends itself to a Stochastic Differential Equation approach \cite[e.g.][]{Gardiner, achterberg1992, MarcowithKirk}, whereby a large number of pseudo-particles are integrated in time, generating a particle distribution that is equivalent to that produced by the Fokker-Planck equation.

Particles are continuously injected at the outer surface of the PWN, at $r=R_{\mathrm{PWN}}(t)$. The region between the PWN and the rest of the SNR interior is assumed to be a uniform reservoir of cosmic rays with energy density
$\langle u_{\rm cr} \rangle=\eta E_{\mathrm{SN}}/{V_\mathrm{Res}}$, where 
$V_{\mathrm{Res}}=\frac{4\pi}{3}(R_{\mathrm{SNR,fs}}(t)^3-R_{\mathrm{PWN}}(t)^3)$ is the reservoir's volume. For simplicity, we take as $R_{\mathrm{SNR,fs}}(t)$ the solution of \citet{mckeetruelove}. Both $V_{\rm res}$ and $\eta$ are in reality time dependent, though since we do not know the history of the external medium in the early stages of the SNR, nor the CR escape history, we take $\eta$ to be a fixed value.

\begin{figure}
        \centering
	\includegraphics[width=0.8\columnwidth]{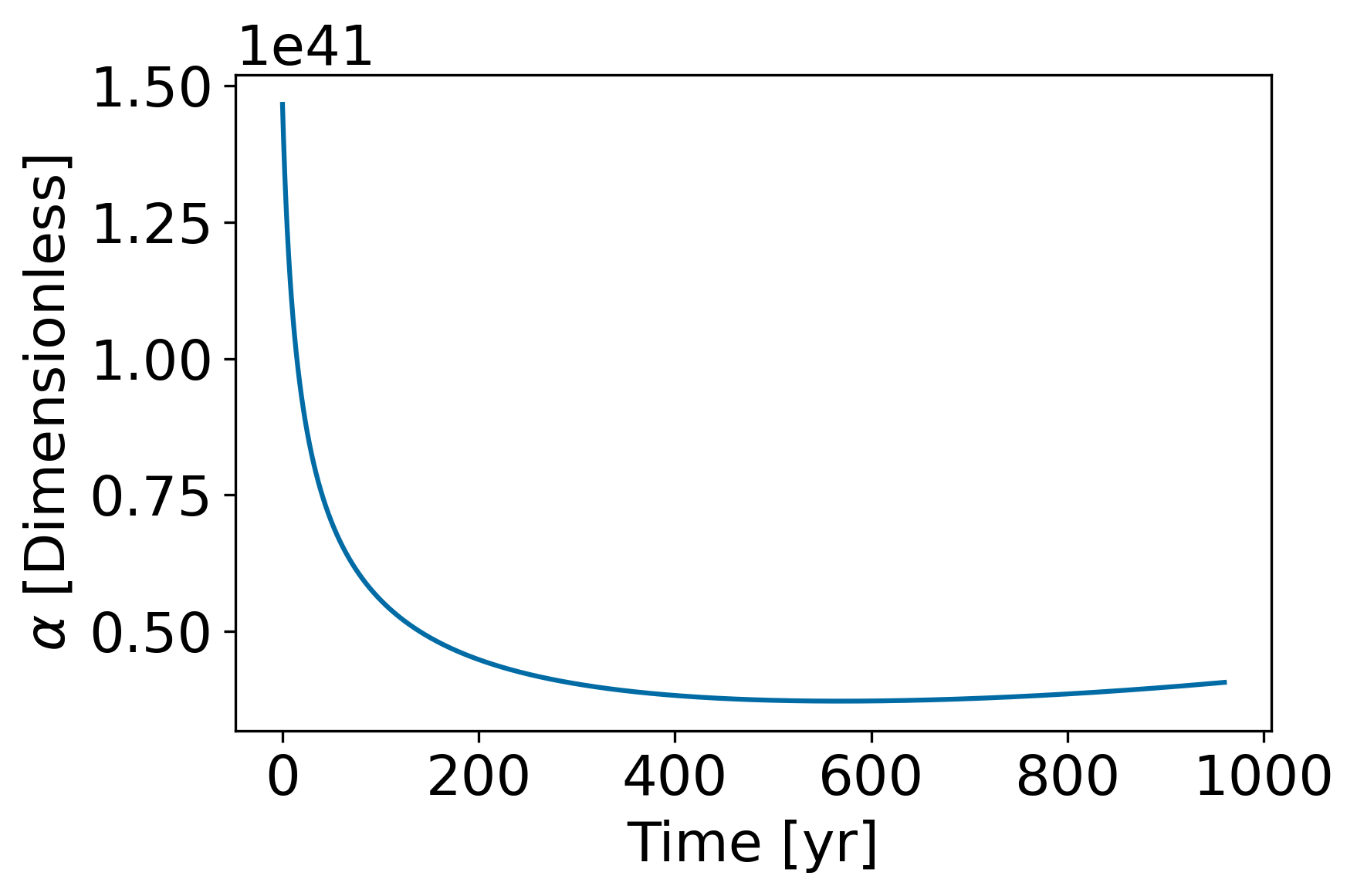}
    \caption{Values of $\alpha$ at injection over the duration of the simulation. $\alpha$ (determined by Equation \ref{eq:alpha}) is assigned to each particle at injection and acts as a weighting factor in the particle spectrum upon escape.}

    \label{fig:alpha}
\end{figure}

The energy distribution of CRs in the reservoir is taken to be a power-law $F \propto (E/E_0)^{-S}$, with $S>1$, where we consider that the particles likely to be re-accelerated have energies between the limits $E_1=1\,\mathrm{TeV}$ and $E_2=50\,\mathrm{TeV}$. Particles are injected at each time step, chosen in such as way that the integrated energy density matches that in the reservoir. This implies a numerical weighting factor for each macro-particle of $\alpha \left(\frac{E}{E_0}\right)^{-S}$, where we inject $N_{\rm inj}$ macro-particles at each timestep, logarithmically spaced between and $\ln(E_1/E_0)$ and $\ln(E_2/E_0)$, such that $\alpha$ is given by
\begin{equation}
\alpha=\frac{4\pi R_{\mathrm{PWN}}^2}{(N_{\rm inj}-1)} (S-1)\ln(E_2/E_1)\left(\frac{\eta E_{\mathrm{SN}}}{E_0}\right)\left(\frac{ v_{\mathrm{PWN}} \Delta t}{\mathrm{V_{res}}(t)}\right)\,.
\label{eq:alpha}
\end{equation}
For the results presented,the number of macro-particles injected per time step is $N_{\rm inj}=2000$. The evolution of $\alpha$ over time is shown in Figure \ref{fig:alpha}.

The evolution of the injected particles are evolved using a standard SDE formalism.
At each timestep, each pseudo-particle's radius $r$ is updated such that 
\begin{equation}
\Delta r = u_{\rm eff}\Delta t + \xi_R \sqrt{2\kappa \Delta t}\,,
\label{eq:deltar}
\end{equation}
where $\xi_R$ is a random number following a standard normal distribution centred at zero with a standard deviation of unity. 

The flow profile in the nebula is chosen to satisfy
\begin{equation}
u_r(r)=\frac{c}{3}\left(\frac{R_{\mathrm{WTS}}}{r}\right)^2,
\end{equation}
following the low magnetisation wind bubble solution \citep{Weaver77} \cite[see also][]{1984ApJ...283..694K}.
Since $u_r$ is divergence free, no change occurs in the particles' energy during any timestep unless the diffusive step results in a pseudo-particle crossing the internal WTS boundary. As stated earlier, the transport equation cannot capture the physics near the shock. We circumvent this issue by simply doubling the particle's energy \cite[as is expected at an ultra-relativistic shock, see][]{achterberg2001} and reflect its position downstream such that $r_{\mathrm{new}}=R_{\mathrm{WTS}}+|R_{\mathrm{WTS}}-r_{\mathrm{old}}|$. Note that for Bohm diffusion, the effective radial velocity for any particle on the shock surface is
\begin{equation}
\frac{u_{\rm eff}}{c} = \frac{1}{3} + \frac{2r_g}{3R_{\mathrm{WTS}}} < 1 \mbox{~~~for~~~} r_g<R_{\mathrm{WTS}}\,.
\end{equation}
Since $r_g = R_{\mathrm{WTS}}$ corresponds to the Hillas limit for relativistic shocks \citep{hillaslimit}, this condition is satisfied. Provided the time step is chosen such that $\sqrt{2\kappa \Delta t} < c\Delta t$, the maximum energy cannot exceed the Hillas limit, since radial outward directed advection must exceed the diffusive step. For our chosen values, the Hillas limit is $E_{\mathrm{Hillas}}\approx 10^{16}$\,eV for protons.

The free parameters used in the simulation are shown in Table \ref{table:params}. The computational domain only covers the region $r_{\rm WTS} < r<r_{\rm PWN}$. Any particle that is transported outside the domain ($r>r_{\rm PWN}$) is recorded, only if it has interacted with the WTS at least once. These particles are injected back into the reservoir and assumed to remain there.

\begin{table*}
\caption{Fixed parameters set in the simulation with associated values based on previous literature.}
\label{table:params}      
\centering   
\resizebox{1.8\columnwidth}{!}{

\begin{tabular}{c c c c}
\hline
Parameter & Description & Value & Notes\\ \hline
$t_0$ & Simulation start time & 9 years & - \\   
$t_{\mathrm{end}}$ & Simulation end time & 969 years & \citet{2003JAHH....6...46S}\\
$B_{\mathrm{Max}}$ & Maximum magnetic field strength in PWN & $\mathrm{112\,\mu G}$ & \citet{lhassocrab}\\
$\Delta t$ & Timestep & 0.01 years & Based on constraints from \citet{LiuWang}.\\
$M_{\mathrm{ej}}$& Mass ejected in supernova & $5 M_{\odot}$ & \citet{1997AJ....113..354F}\\
$E_{\mathrm{SN}}$ & Supernova energy & $\mathrm{10^{51}\,erg}$ & -\\
$\eta$ & Fraction of $E_{\mathrm{SN}}$ in protons in the reservoir & $7\times 10^{-5}$ & -\\
$E_0$ & Proton normalisation energy & $\mathrm{1\,GeV}$ & -\\
$\dot{E}$ & Spin-down luminosity of Crab pulsar at current time& $\mathrm{5\times 10^{38}\,erg\,s^{-1}}$& \citet{hester}\\
$n_{\mathrm{ISM}}$ & Proton density in ISM & $\mathrm{0.1\,cm^{-3}}$ & -\\
$n_{\mathrm{Target}}$ & Proton density in target material & $\mathrm{5\,cm^{-3}}$ & - \\
$D$ & Distance to Crab nebula & $\mathrm{2\,kpc}$ & \citet{Kaplan_2008} \\
$\beta$ & Diffusion coefficient relative to Bohm diffusion & 1 & -\\
$N_{\mathrm{inj}}$ & Pseudo-particles injected per timestep & 2000 & - \\
$E_1$ & Minimum pseudo-particle injection energy & $\mathrm{1\,TeV}$ & - \\
$E_2$ & Maximum pseudo-particle injection energy & $\mathrm{50\,TeV}$ & - \\
$S$ & Pseudo-particle injection spectral index & 1.2 & - \\

\hline
\end{tabular}
}
\end{table*}

\subsection{$\gamma$-ray spectrum}
The package GAMERA \citep{GAMERA,2022ascl.soft03007H} is used to infer the $\gamma$-ray spectrum from the protons interacting with the reservoir using the cross-section parameterisations of \citet{2014PhRvD..90l3014K}. The reservoir is taken as a static target with a $5\,\mathrm{cm}^{-3}$ density, with the proton spectra from the simulation at 969 years being used as the input proton distribution. It should be noted that the normalisation of the hadronic spectrum here is effectively a free parameter given the assumptions about the density in the target material and the fraction of the supernova energy in the population of re-accelerated hadrons. To model the hadronic emission, we employed the SYBIL 2.1 \citep{sibyll} code as supported by GAMERA.

\section{Results and discussion}
\label{sec:results}

\subsection{Comparison to $\gamma$-ray data}
\begin{figure}
        \centering
	\includegraphics[width=\columnwidth]{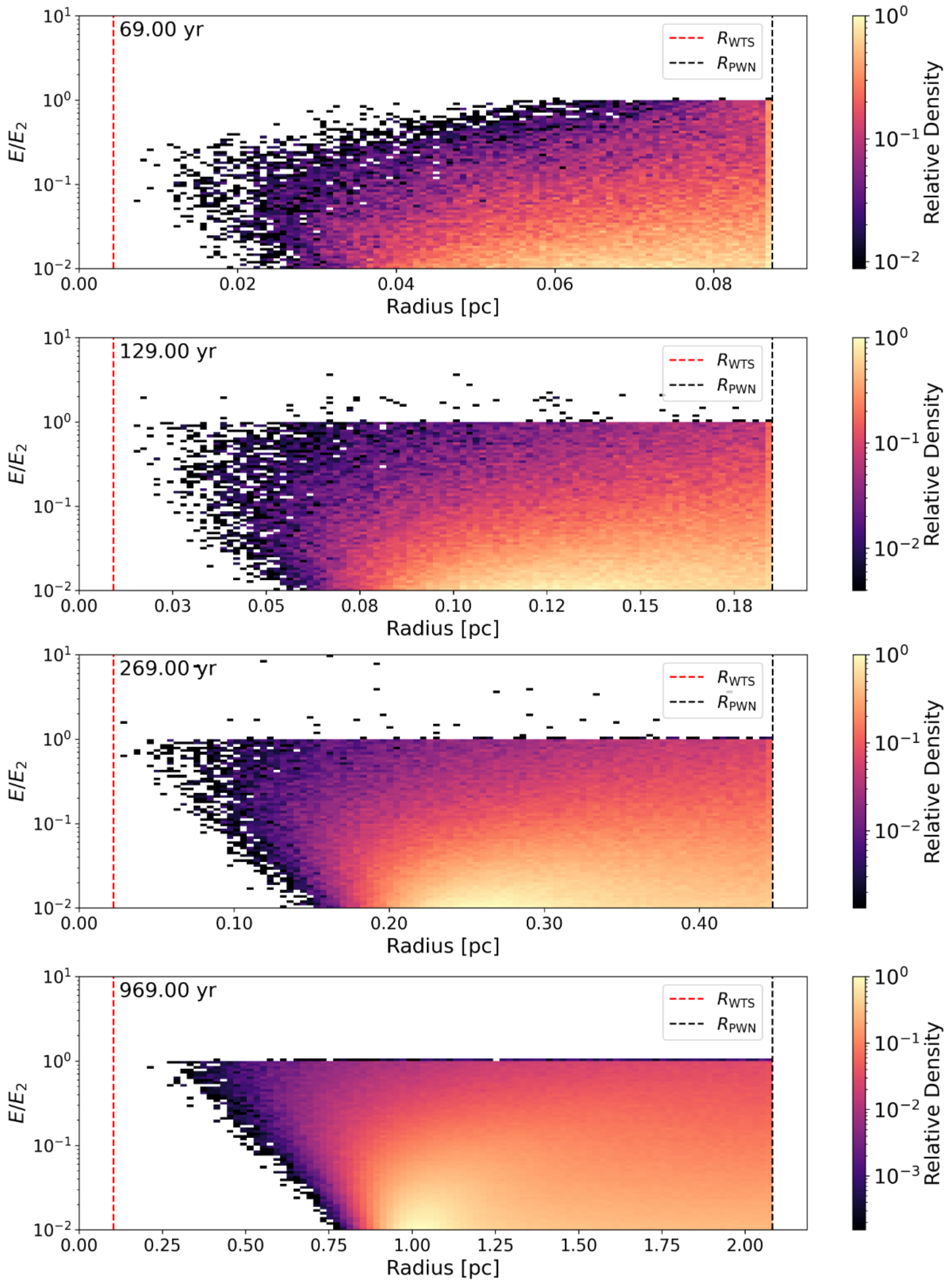}
    \caption{Instantaneous steps of the simulation over time. At each time step, the distribution of particle energies compared to the maximum injection particle energy is shown as function of distance from the pulsar. A small number of particles with energy ratios greater than 1 can be seen, corresponding to interactions with the WTS. The PWN radius is constrained to reach the current size at the end of the simulation, 969 yrs. At late times ($\gtrsim 650$ years), there are no interactions with the WTS and very few shocked particles remain within the nebula.}

    \label{fig:multiplot}
\end{figure}

\begin{figure}[h]
    \centering
    \includegraphics[width=0.9\columnwidth]{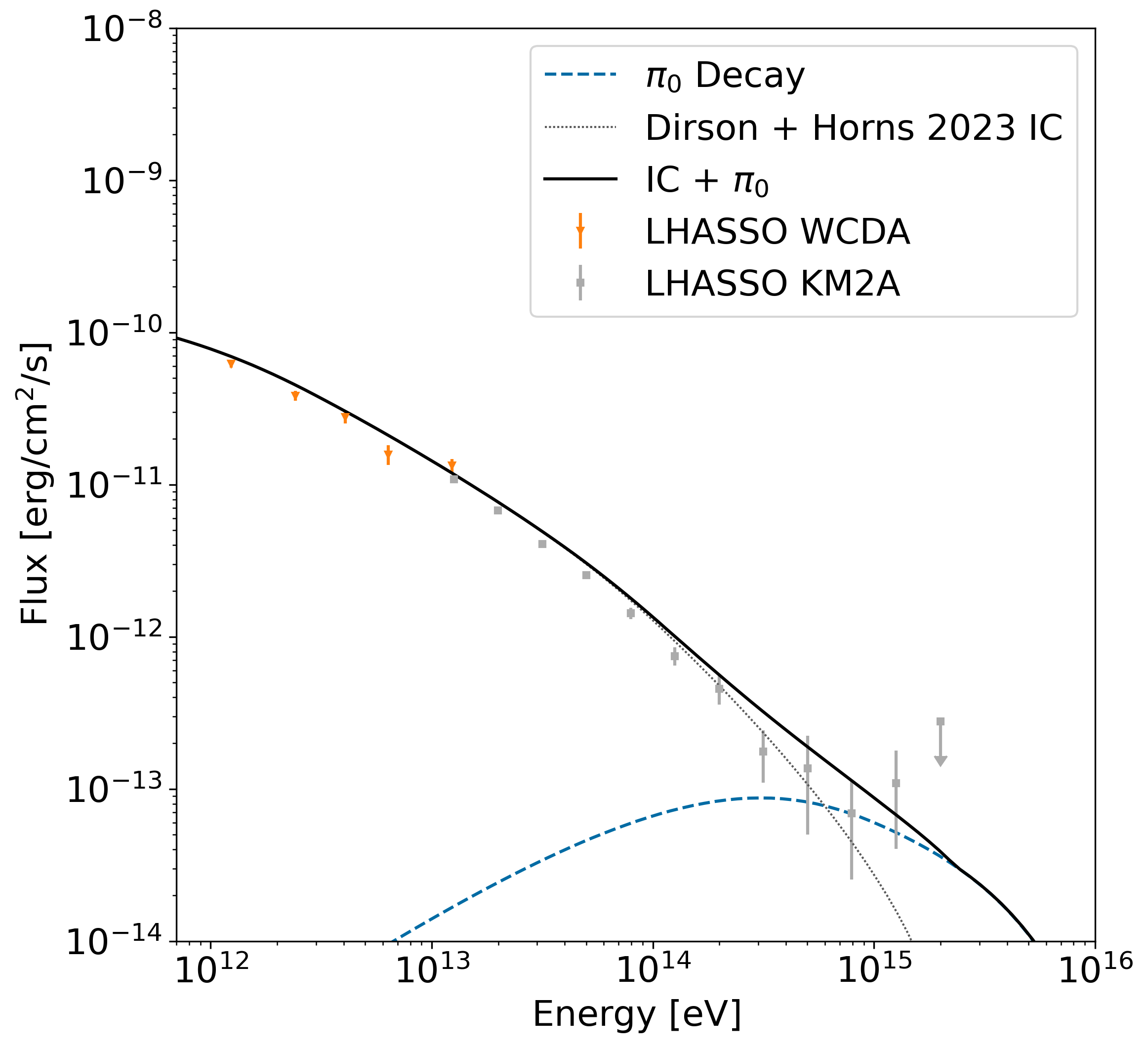}
    \caption{Hadronic Crab spectra as modelled using GAMERA with SIBYLL 2.1. The Inverse Compton spectra from \citet{dirsonhorns} is also shown.}
    \label{fig:crabspec}
\end{figure}

\begin{figure}
        \centering
	\includegraphics[width=0.8\columnwidth]{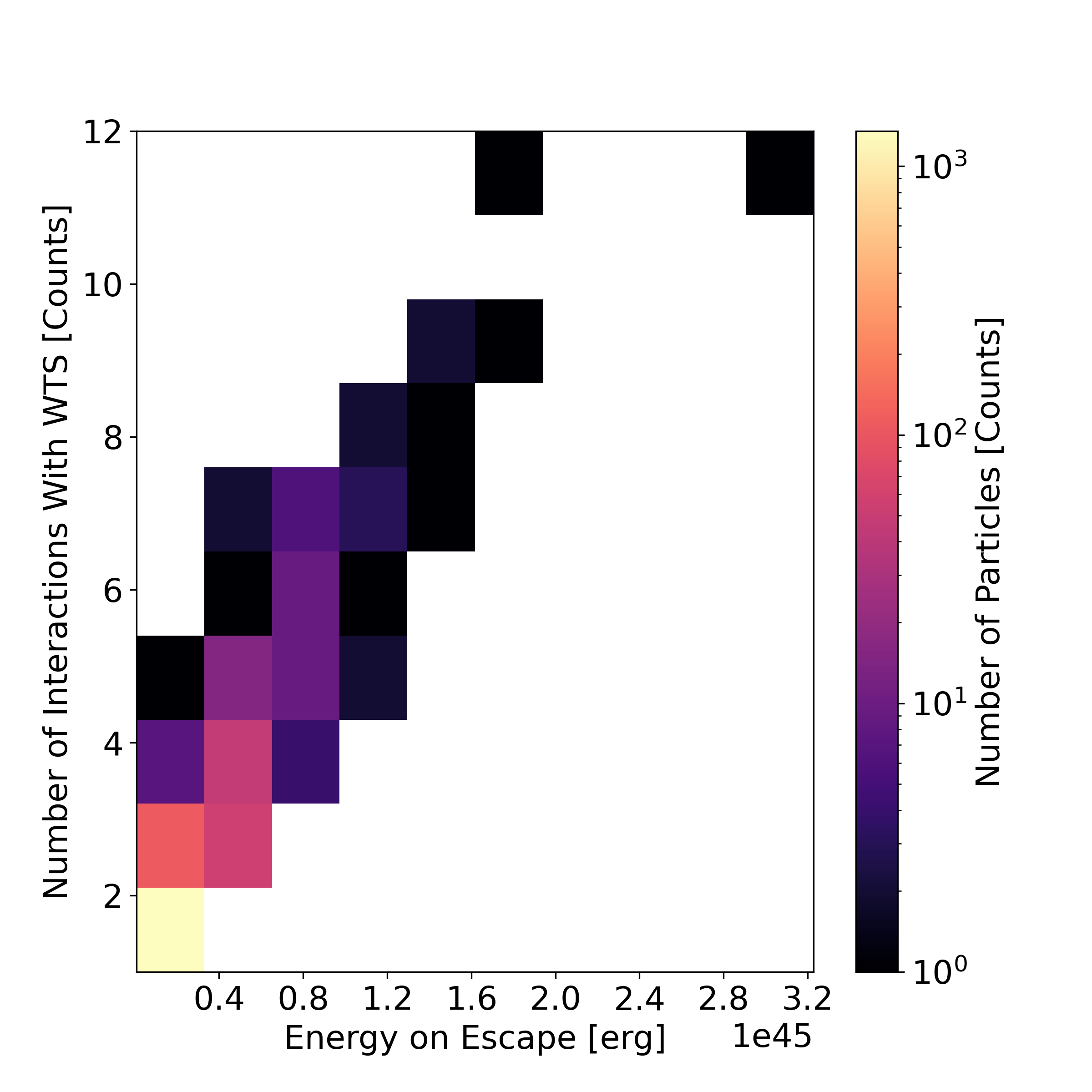}
    \caption{Re-weighted particle energies on escape from the PWN as a function of the number of interactions with the WTS. Only a small number of the $\sim10^8$ total particles injected over the PWN's history ever reach the WTS, of which a few 10s can interact with the shock $\gtrsim10$ times.}
    \label{fig:shockhistalpha}
\end{figure}

\begin{figure}[h]
        \centering
	\includegraphics[width=0.8\columnwidth]{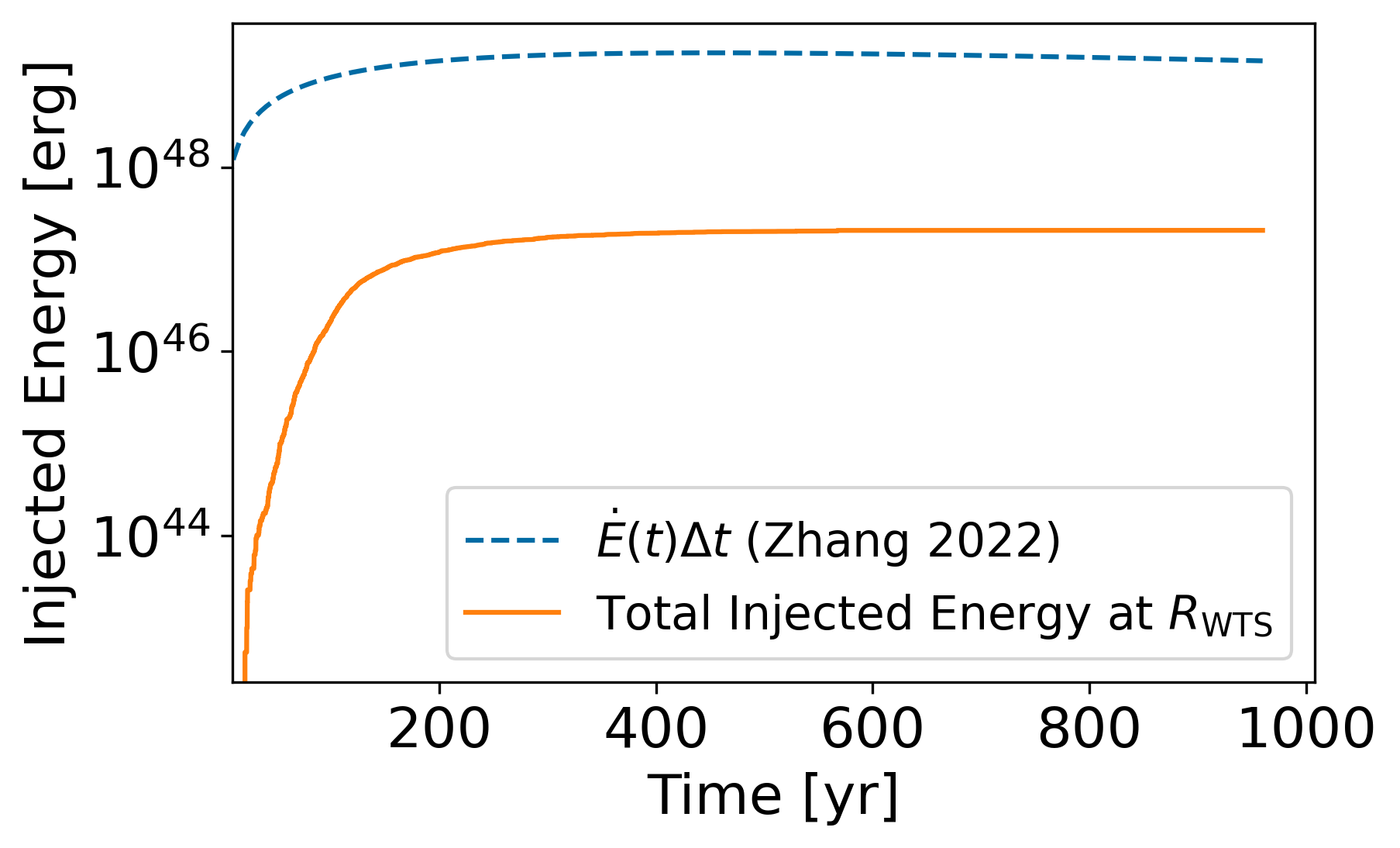}
    \caption{The spin-down energy output of the pulsar compared to the total energy injected by the pulsar into re-accelerated hadrons at the WTS.}
    \label{fig:en}
\end{figure}

Figure \ref{fig:multiplot} shows instantaneous steps of our simulations at selected times, which particle energy relative to the maximum injection energy as a function of radius from the pulsar. This shows that the majority of re-accelerated particles interact with the WTS comparatively early (with the first interactions occurring around 30 years), before advecting outwards with the wind. The majority of interactions occur at approximately $\sim100$ years. After $\sim650$ years the simulations effectively freeze-out, with no high energy particles being shocked or remaining within the PWN, as expected. These results further indicate that protons re-accelerated at the WTS are, as anticipated, outliers. If the contrary were true, their emission would potentially exceed the observed flux at lower energies.
A secondary population of electrons and positrons produced through the secondary hadronic population's proton-proton interactions would not be detectable above the existing inverse Compton emission from the primary electrons.
\begin{figure}[h]
        \centering
	\includegraphics[width=0.8\columnwidth]{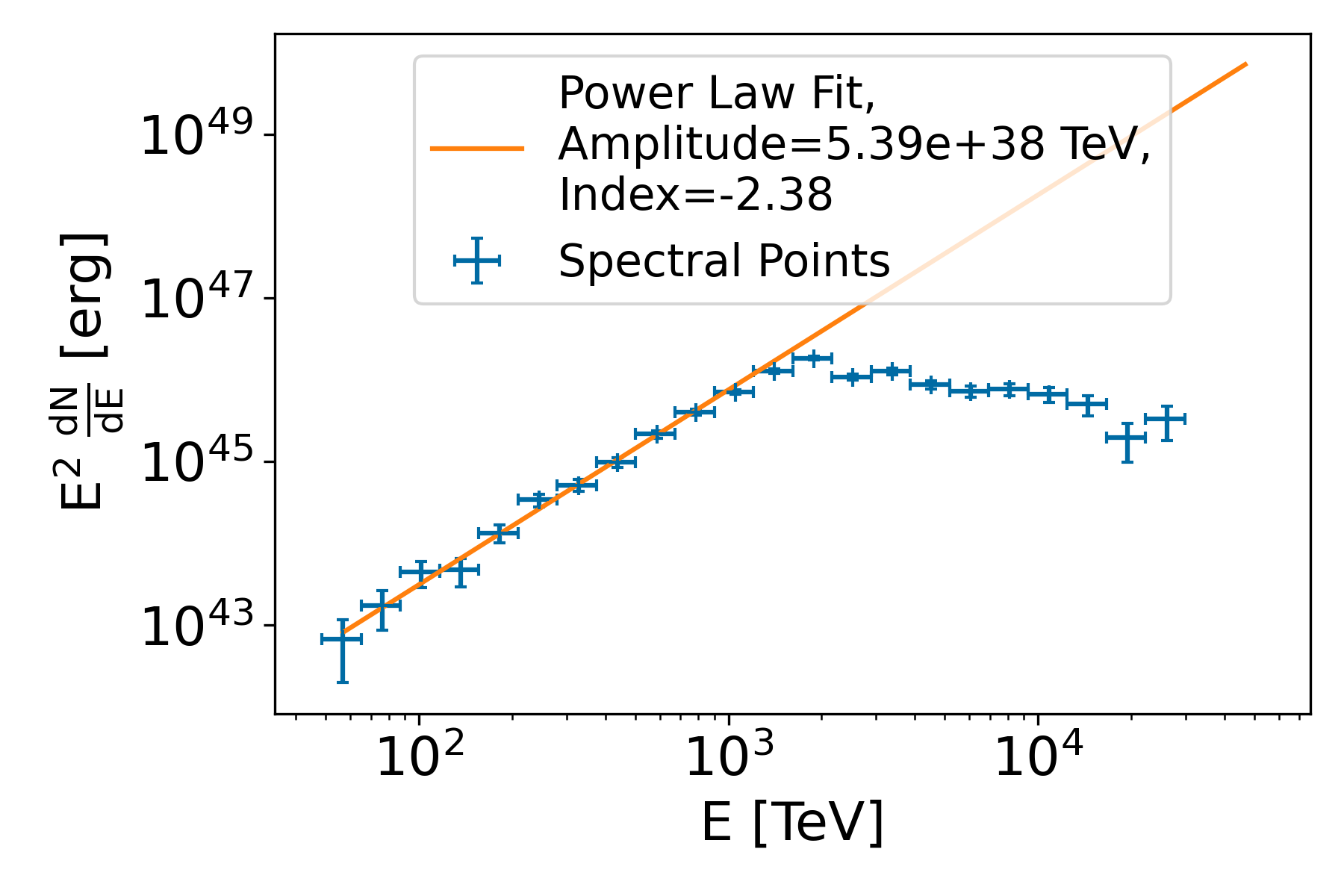}
    \caption{Spectrum of re-accelerated protons that have escaped the PWN at the final timestep, the Poisson errors on these measurements are shown, the $x$-axis errors denote the size of the bins. A power-law fit of $E^2 dN/dE$ up to $1\,\mathrm{PeV}$, normalised to $1\,\mathrm{TeV}$, is shown for comparison to highlight the spectral break.}

    \label{fig:reacceleratedspec}
\end{figure}
Figure \ref{fig:reacceleratedspec} shows the spectrum of re-accelerated protons that exit the PWN and merge with the CR reservoir; a clear break at $\sim\mathrm{1\,PeV}$ is seen. The resulting $\gamma$-ray spectrum is found using GAMERA, and is compared to the LHAASO flux points in Figure \ref{fig:crabspec}. Here the multi-band leptonic model fit from \citet{dirsonhorns} is taken for the IC component. Our re-acceleration model provides a reasonable match to the data, indicating that a hadronic re-acceleration scenario cannot be excluded as the origin of the highest energy emission. A longer integration time with LHAASO or next-generation detectors such as the Cherenkov Telescope Array Observatory (CTAO) will help to constrain this hadronic scenario further. In particular, the improved angular resolution of CTAO ($\sim0.02^{\circ}$ at $100\,\mathrm{TeV}$) could help determine if the highest energy $\gamma$-rays are correlated with the position of the target gas that sits outside the nebula \citep{cherenkov_telescope_array_observatory_2021_5499840}.
A very small number of particles ($\sim10$s out of $\sim10^8$ injected) interacting with the shock $>10$ times appears to be able to account for the highest energy emission, this can be seen in Figure \ref{fig:shockhistalpha}. These tend to be the particles injected with lower energies as a result of the statistics of the injection.
To confirm that the pulsar can provide sufficient power to re-accelerate the protons, we compare the energy gained by the pseudo-particles at the WTS to the pulsar spin-down luminosity of the pulsar $\dot{E}(t)$ over time
\begin{equation}
\dot{E}(t)=\dot{E}_0\left(1+\frac{t}{\tau_0}\right)^{-\frac{(n+1)}{(n-1)}},
\end{equation}
where $\dot{E}_0$ is the spin-down luminosity at birth and $n$ is the braking index (which we fix to be 2). The characteristic age of the pulsar $\tau_0$ is given by
\begin{equation}
\tau_0=\frac{P_0}{(n-1)\dot{P}_0}
\end{equation}
where we take the values from \citet{zhang} for the Crab pulsar birth period $P_0=18.3\,\mathrm{ms}$ and birth period derivative $\dot{P}_0=6.4\times10^{-13}\,\mathrm{s\,s^{-1}}$. To calculate $\dot{E}_0$ we use the equation
\begin{equation}
\dot{E}_0=4\pi^2I\frac{\dot{P}_0}{P_0^3}
\end{equation}
\cite[c.f.][]{Gaensler_2006}, where $I=10^{45}\,\mathrm{g\,cm^2}$ is the neutron star's moment of inertia. The results of this comparison are shown in Figure \ref{fig:en}. In order to match the highest energy LHAASO flux point, we select $\eta$ to be $7\times 10^{-5}$. This suggests that in our scenario only a small portion of the pulsar's spin-down luminosity would ultimately go into the re-acceleration of CRs. We also performed simulations with $S$ in the range 1.1-1.3, and with $E_2$ as 25\,TeV and 75\,TeV; the effect on the spectrum was minor and would not change our conclusions, especially given the freedom of the selected value of $\eta$.
\subsection{Predicted neutrino flux}

\begin{figure}[h]
\centering
	\includegraphics[width=0.8\columnwidth]{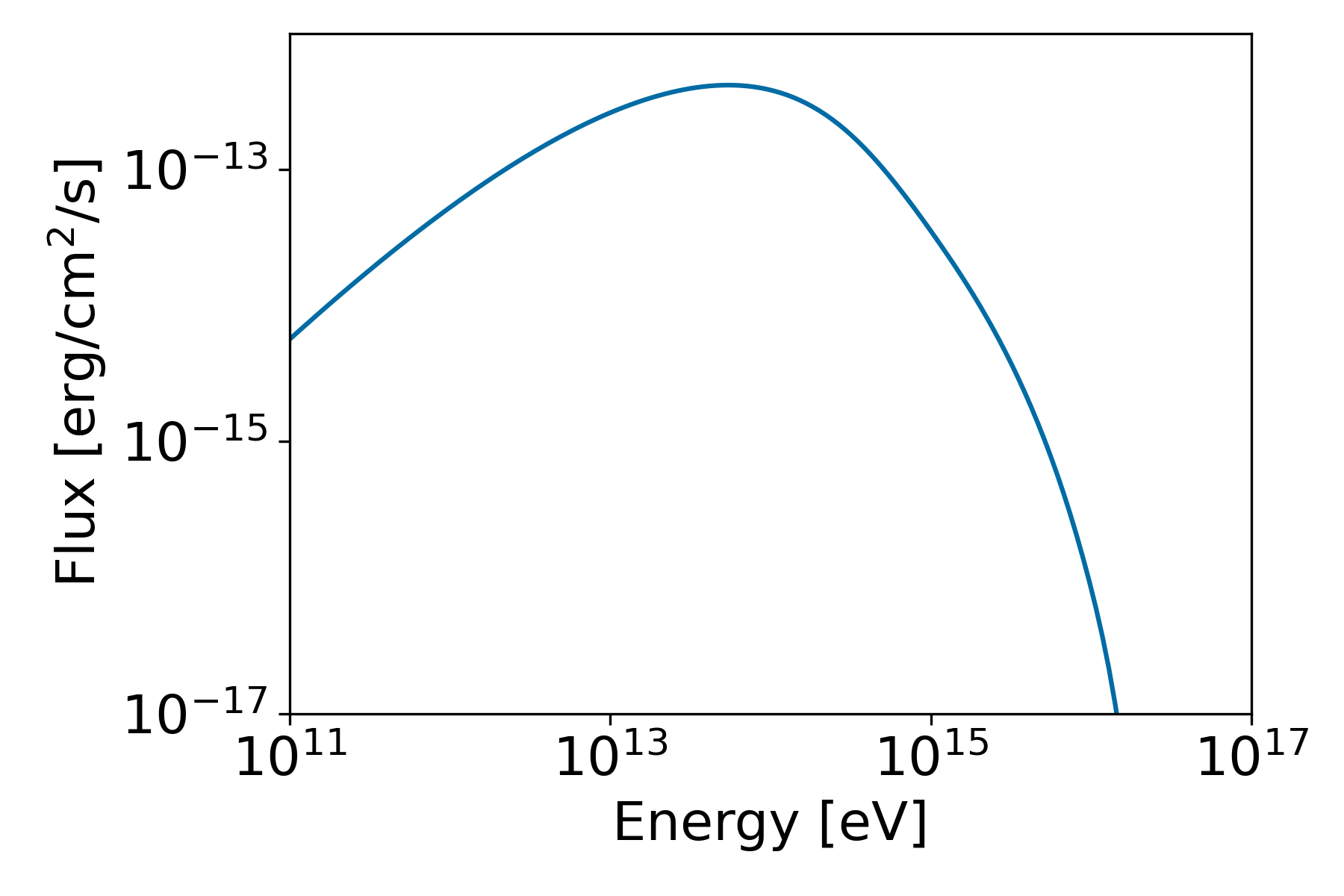}
    \caption{Neutrino flux prediction for our re-acceleration model, for reference the current generation of neutrino detectors have an optimal sensitivity at $100\,\mathrm{TeV}$ of approximately $10^{-12}\,\mathrm{erg\,cm^{-2}\,s^{-1}}$ \citep{icecubesens}.}
    \label{fig:nu}
\end{figure}

The neutrino flux we predict from our secondary hadronic population, according to the prescription of \citet{kelner2006} is shown in Figure \ref{fig:nu}. Similar to \citet{peng_multiband_2022}, we find that the predicted neutrino flux is substantially below the sensitivity threshold of current generation instruments such as the IceCube neutrino observatory \citep{icecube} and the upcoming KM3Net neutrino detector \citep{km3net}. This could potentially be in tension with the hotspot associated with the Crab in the recent cascade event analysis performed by the IceCube collaboration \cite{icecube}.

\section{Conclusions}
\label{sec:conclusions}
In this work we have performed particle transport simulations to discern whether the highest energy emission from the Crab Nebula as observed by LHAASO could have a hadronic contribution, assuming a physically motivated re-acceleration scenario. Our work shows that, in principle, protons have sufficient time to travel from the region between the PWN radius and the SNR forward shock to the WTS, interact multiple times with the WTS region, and travel back again. Our model provides a reasonable quality match to the LHAASO data without over-estimating the flux at lower energies, indicating that a hadronic scenario cannot be excluded. Our results motivate further, more realistic 3D simulations to continue exploring this scenario. Possible extensions of this model could take into account the reverberation phase in older systems, in which the combination of crushing of the PWN by the SNR reverse shock \citep{ohira} and re-acceleration at the central WTS could lead to PeV particles being produced in greater numbers.

\begin{acknowledgements}
S. Spencer and A. Mitchell are supported by the Deutsche Forschungsgemeinschaft (DFG, German Research Foundation) – Project Number 452934793.
\end{acknowledgements}

\bibliographystyle{aa}
\bibliography{example}
\end{document}